\documentclass[aps,twocolumn,prb]{revtex4}

\usepackage{graphicx,psfrag,epsfig,amsfonts}

\begin{document}
\title{
Finite size effects, super- and sub-poissonian noise in a nanotube connected to leads
}

\author{Marine Guigou$^a$}
\author{Alexandre Popoff$^a$}
\author{Thierry Martin$^a$}
\author{Adeline Cr\'epieux$^a$}

\affiliation{
$^a$Centre de Physique Th\'eorique, Universit\'e de la
M\'editerran\'ee, Case 907, 13288 Marseille, France}

\begin{abstract} 
The injection of electrons in the bulk of carbon nanotube which is connected 
to ideal Fermi liquid leads is considered. While the presence of the leads 
gives a cancellation of the noise cross-correlations, the auto-correlation
noise has a Fano factor which deviates strongly from the Schottky behavior
at voltages where finite size effects are expected. Indeed, as the voltage is increased from 
zero, the noise is first super-poissonian, then sub-poissonian, and eventually it 
reaches the Schottky limit. These finite size effects are also tested using 
a diagnosis of photo-assisted transport, where a small AC modulation is superposed 
to the DC bias voltage between the injection tip and the nanotube. When finite 
size effects are at play, we obtain a 
stepwise behavior for the noise derivative, as expected for normal metal systems, whereas in the absence of finite size effects, due to the presence of Coulomb interactions, a smoothed staircase is observed. The present work shows that it is possible to explore finite size effects in nanotube transport
via a zero-frequency noise measurement.
\end{abstract}

\maketitle

%
%
%
%

\section{Introduction}

An important issue for transport in nanosystems concerns the role of electronic interactions. For one dimensional conductors such as carbon nanotubes, electronic correlations are known to lead to dramatic behavior, such as a zero bias anomaly 
in the tunneling conductance\cite{kane_balents_fisher,mceuen}. Recently, the problem of electron injection in the bulk of a nanotube, which extremities are connected to leads, was examined\cite{crepieux}. 
For an infinite carbon nanotube length, it was shown that the measurement of both current auto- and cross-correlations in the current could lead to a diagnosis of the anomalous (non-electron) charges arising from collective excitations propagating in the nanotube.  This work was followed other studies\cite{trauzettel,lebedev} where the role of Fermi liquid leads connected to the nanotube was investigated in view of detecting anomalous charges. 
For the injection geometry of Ref.~\onlinecite{lebedev}, 
the zero-frequency noise cross-correlations vanish as a result of multiple Andreev-like reflections at the contacts: only electrons can be accepted by the right and left contacts. To a first approximation, current and 
auto-correlation noise were shown to follow the Schottky relation\cite{schottky} with a Fano factor corresponding to the electron charge. The purpose of the present work is 
to focus on the finite size effects which are manifest in this transport 
geometry, and is two-fold. 

First, we wish to reexamine whether this Schottky relation is indeed
followed for all parameters in this finite size geometry. Indeed, there are several frequency scales in this geometry. The inverse of the voltage scale $\omega_0=eV_0/\hbar$ corresponds to the time spread of the electron wave packet 
entering the nanotube. The inverse of the finite length frequency 
$\omega_L=2v_F/K_{c+}L$ ($K_{c+}$ is the interaction parameter of the nanotube, $L$ is its length) corresponds to the time of flight of excitations propagating from one contact to the other. We shall show that in the limit where $\omega_0<\omega_L$, the Schottky behavior -- with the electron charge as the proportionality factor -- is violated, 
leading to a voltage dependent Fano factor.

The second part of this study concerns photo-assisted noise: an AC bias is superposed to the DC bias imposed between the Scanning Tunneling Microscope~(STM) 
tip which injects the electrons, and the  nanotube. Experimentally, photo-assisted noise has been measured in diffusive wires\cite{schoelkopf}, diffusive junctions\cite{kozhevnikov} and quantum point contacts\cite{reydellet}. For normal metals, the noise 
derivative displays steps at integer values of the ratio $\omega_0/\omega$,
where $\omega$ is the AC frequency\cite{lesovik}. We naturally expect that this behavior
is modified by the tunneling density of states exponent of our geometry, leading 
here to a smoothing of the steps due to electronic correlations. It is the case in the absence of finite size effects. More interestingly, we will show that when finite size effects are present, the noise derivative with respect to voltage has a stepwise behavior, similar to the one obtained for normal metal.

Indeed, concerning the first part of this study, our previous works on electron injection\cite{crepieux,lebedev} did not fully take into account the presence of the leads, in the sense
that the tunneling density of states is slightly modified by finite size effects.
Note that this does not bear any dramatic consequences on our finite frequency 
noise scheme \cite{lebedev} which was used to detect anomalous 
charges in a nanotube connected to Fermi liquid leads. In the present zero-frequency noise study, we feel that it is quite interesting to deepen the study 
of transport because to our knowledge, in most tunneling geometries of mesoscopic physics, the Schottky relation is expected to be followed with a voltage
independent Fano factor. The present numerical study allows to probe to what 
extends this relation is violated when the voltage is ``low'', and this violation 
is rendered more explicit when interactions in the nanotube are strong.

The second part of this study deals with photo-assisted noise. Photo-assisted transport (current) is by no means new: a pioneering work\cite{tien} considered the effect of a microwave fields on the tunnel transfer of electrons between two superconductor films. In mesoscopic devices, photo-assisted noise was first studied theoretically in a normal metal junction, leading, as mentioned above, to a stepwise behavior in the nose derivative\cite{lesovik}. This latter work was extended to treat normal metal-superconductor junctions\cite{lesovik_martin_torres}, leading to a 
diagnosis of the Cooper pair charge transferred in an Andreev reflection 
process. It was also applied to the fractional quantum Hall 
effect\cite{crepieux2} where the charge
transferred can either be that of a Laughlin quasiparticle or that of an electron, 
depending on whether weak or strong pinching of the point contact placed 
on a quantum Hall bar.
So as far as 1D strongly correlated systems are concerned, photo-assisted transport has so far been confined only to chiral Luttinger liquids.
A further study of photo-assisted transport in non-chiral Luttinger liquids is clearly lacking, but most importantly its crucial feature here is to understand 
how finite size effects of the nanotube modify the ``expected'' behavior.

The paper is organized as follows: in Sec. II, we define the geometry which applies to our calculation and we recall the basic assumptions of the model
developed in Ref. \onlinecite{lebedev}. In Sec. III, we 
concentrate on the calculation of the nanotube current and of the nanotube auto-correlation noise for the numerical study of the voltage dependent
Fano factor. Sec. IV, is devoted to the calculation of photo-assisted 
transport in a nanotube. We conclude in Sec. V.     

%
%
%
%

\section{Model}

\begin{figure}[h]
\epsfxsize 8 cm
\centerline{\epsffile{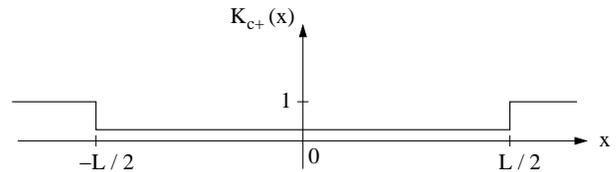}}
\caption{For the total charge sector, we assume that the Coulomb interactions parameter $K_{c+}$ is equal to 1 in the leads (i.e., for $|x|\ge L/2$) and is smaller than 1 in the nanotube (i.e., for $|x|< L/2$) due to Coulomb interactions.\label{fig1}}
\end{figure}

We consider the following setup: an STM tip close to a carbon nanotube connected to leads at both extremities. A voltage applied between the STM and the nanotube allows electrons to tunnel in the center region of the nanotube. As a result, charge excitations propagate along the nanotube toward the right and left leads. This system is described by the Hamiltonian $H=H_\mathrm{N}+H_\mathrm{STM}+H_\mathrm{T}$. The nanotube is a non-chiral Luttinger liquid\cite{egger_98}:
\begin{eqnarray}
      H_\mathrm{N}&=&\frac12\sum\limits_{j\delta}\int\limits_{-\infty}^{+\infty}
      dx\;v_{j\delta}(x)\Bigl[K_{j\delta}(x)(\partial_x\phi_{j\delta})^2
      \nonumber\\
      &\ &\qquad+
      K_{j\delta}^{-1}(x)(\partial_x\theta_{j\delta})^2\Bigr]~,
      \label{b_ham}
\end{eqnarray}
where $x$ is the position along the nanotube, $\phi_{j\delta}$ and $\theta_{j\delta}$ are non-chiral bosonic fields and $K_{j\delta}$ is the Coulomb interactions parameter for each charge/spin, total/relative sectors $j\delta\in\{c+,c-,s+,s-\}$. We put $\hbar=1$. Because of time reversal 
symmetry, $K_{c-}(x)=K_{s+}(x)=K_{s-}(x)=1$, and we assume that $K_{c+}$ depends on position\cite{safi_maslov} as depicted on Fig.~\ref{fig1}. The velocities satisfy $v_{j\delta}(x)=v_\mathrm{F}/K_{j\delta}(x)$.

The electrons in the metallic STM tip are assumed to be non-interacting. For convenience\cite{crepieux}, the electron field $c_\sigma(t)$ in the STM tip can be described in terms of a semi-infinite Luttinger liquid with Coulomb interactions parameters all equal to one. The tunnel Hamiltonian between the STM tip and the nanotube at position $x=0$ is:
\begin{eqnarray}
      H_\mathrm{T}(t)=\sum\limits_{r\alpha\sigma\epsilon}
      \Gamma^{(\epsilon)}(t)\bigl[\Psi_{r\alpha\sigma}^\dagger(0,t)
      c_\sigma(t)
      \bigr]^{(\epsilon)}~,
      \label{ham}
\end{eqnarray}

where $r$ corresponds to the branch index, $\alpha$ to the mode index and $\sigma$ to the spin. 
 The superscript
$(\epsilon)$ leaves either operator unchanged $\epsilon=+$, or
transforms it into its hermitian conjugate $\epsilon=-$. 
The voltage is taken into account via a time dependence
of the tunneling amplitude (Peierls substitution)
$\Gamma(t)=\Gamma\,e^{i\omega_0 t}$ where
$\omega_0=eV_0/\hbar$ is the voltage frequency.
The fermionic fields for electrons in the nanotube and in 
the STM tip are respectively defined by:
\begin{eqnarray}
      \Psi_{r\alpha\sigma}(x,t)&=&\frac{F_{r\alpha\sigma}}{
      \sqrt{2\pi a}}\,e^{ik_{\rm\scriptscriptstyle F}rx+
      iq_{\rm\scriptscriptstyle F}\alpha x+i\varphi_{r\alpha\sigma}
      (x,t)}~,\\
      c_\sigma(t)&=&\frac{f_\sigma}{\sqrt{2\pi a}}\,e^{i\tilde
      \varphi_\sigma(t)}~,
\end{eqnarray}
where $a$ is the ultraviolet cutoff of the Luttinger liquid
model, $F_{r\alpha\sigma}$ and $f_\sigma$ are Klein factors,
$k_{\scriptscriptstyle\rm F}$ is the Fermi momentum and
$q_{\scriptscriptstyle\rm F}$ is the momentum mismatch associated
with the two modes $\alpha$. For further calculation purposes, it is
convenient to rewrite the bosonic field
$\varphi_{r\alpha\sigma}$ in terms of the non-chiral bosonic
fields $\theta_{j\delta}$ and $\phi_{j\delta}$:
\begin{equation}
      \varphi_{r\alpha\sigma}(x,t)=\frac{\sqrt{\pi}}2
      \sum\limits_{j\delta}h_{\alpha\sigma j\delta}\,
      \bigl[\phi_{j\delta}(x,t)+r\theta_{j\delta}(x,t)
      \bigr]~,
\end{equation}
with coefficients $h_{\alpha\sigma c+}=1$, $h_{\alpha\sigma
c-}=\alpha$, $h_{\alpha\sigma s+}=\sigma$ and $h_{\alpha\sigma
s-}=\alpha\sigma$, and bosonic fields obeying the equal time
commutation relations
$[\phi_{j\delta}(x),\theta_{j^\prime\delta^\prime}(x^\prime)]=-(i/2)
\delta_{jj^\prime}\delta_{\delta\delta^\prime}\,sgn(x-x^\prime)$.
$\tilde \varphi_\sigma(t)$ is the chiral bosonic field attached to the STM tip,
whose Keldysh Green's function at $x=0$ is given
by~\cite{chamon_99}:
\begin{eqnarray}\label{green}
      &&g_{\sigma(\eta\mu)}(t_1-t_2)=\Bigl\langle
      T_\mathrm{K}\bigl\{\tilde\varphi_\sigma(t_1^{\eta})
      \tilde\varphi_\sigma(t_2^{\mu})\bigr\}\Bigr\rangle
      \label{STM_gr}\\
      &&=\!-\!\ln\!\left[1+i(\eta\!+\!\mu)
      \frac{v_{\scriptscriptstyle\rm F}
      |t_1\!-\!t_2|}{2a}-i(\eta\!-\!\mu)
      \frac{v_{\scriptscriptstyle\rm F}
      (t_1\!-\!t_2)}{2a}\right],\nonumber
\end{eqnarray}
where $\eta,\mu=\pm1$ denotes the upper/lower branch of the
Keldysh contour.

%
%
%
%

\section{Current and noise in the nanotube}

In this section, we present the calculations of current and zero-frequency shot noise. Since we have to treat a non-equilibrium situation due to the application of a voltage bias between the STM tip and the nanotube, we define average values of the current and of the 
unsymetrized noise in the framework of the Keldysh formalism\cite{keldysh}:
\begin{eqnarray} \label{courant_moyen}
\langle I(x,t)\rangle&=&\frac{1}{2}\sum_\eta\langle T_K\{\hat I(x,t^\eta)
e^{-i\int_K dt_1H_T(t_1)}\}\rangle~,\\
S(x,x',t,t')&=&\left\langle T_\mathrm{K}
\Bigl\{\hat I(x,t^-)\hat I(x^\prime,t^{\prime+})
e^{-i\int_\mathrm{K} dt_1H_\mathrm{T}(t_1)}\Bigr\}
\right\rangle~,\nonumber\\
\end{eqnarray}
where $T_K$ denotes time ordering along the Keldysh contour and $\hat I(x,t)$ is the total current operator which can be defined through the bosonic field $\phi_{c+}$: $ \hat I(x,t)=2ev_{\scriptscriptstyle\rm F}\,\partial_x\phi_{c+}(x,t)/\sqrt\pi$. It has been shown in Ref.~\onlinecite{crepieux} that, up to the second order with the tunnel amplitude $\Gamma$, the average current and the zero-frequency Fourier transform of the noise can be expressed as:
\begin{widetext}
\begin{eqnarray} \label{courant}
&&\langle I(x)\rangle=
-\frac{ev_F\Gamma^2}{2\pi^2 a^2}\sum_{\eta\eta_1r_1\sigma_1}\nonumber\\
&&\times\int_{-\infty}^{+\infty}d\tau'\partial_x\left(G^{\phi\phi}_{c+(\eta\eta_1)}(x,0,\tau')
-G^{\phi\phi}_{c+(\eta-\eta_1)}(x,0,\tau')+r_1G^{\phi\theta}_{c+(\eta\eta_1)}(x,0,\tau')
-r_1G^{\phi\theta}_{c+(\eta-\eta_1)}(x,0,\tau')\right)\nonumber\\
&&\times \int_{-\infty}^{+\infty}d\tau sin(\omega_0\tau)e^{2\pi
g_{\sigma_1 (\eta_1-\eta_1)}(\tau)}
e^{\frac{\pi}{2}\sum_{j\delta}(G^{\phi\phi}_{{j\delta}(\eta_1-\eta_1)}(0,0,\tau)
+G^{\theta\theta}_{{j\delta}(\eta_1-\eta_1)}(0,0,\tau)
+r_1G^{\phi\theta}_{{j\delta}(\eta_1-\eta_1)}(0,0,\tau)
+r_1G^{\theta\phi}_{{j\delta}(\eta_1-\eta_1)}(0,0,\tau))}~,
\label{general_nanotube_current}\nonumber\\
\\
&&S(x,x',\Omega=0)=-\frac{e^2v^2_F\Gamma^2}{(\pi a)^2}\sum_{\eta\eta_1r_1\sigma_1}\nonumber\\ \label{bruit}
&&\times \int_{-\infty}^{+\infty}d\tau cos(\omega_0\tau)
e^{2\pi g_{\sigma_1 (\eta_1-\eta_1)}(\tau)}
e^{\frac{\pi}{2}\sum_{j\delta}
(G^{\phi\phi}_{j\delta(\eta_1-\eta_1)}(0,0,\tau)+r_1G^{\phi\theta}_{j\delta(\eta_1-\eta_1)}(0,0,\tau)
+r_1G^{\theta\phi}_{j\delta(\eta_1-\eta_1)}(0,0,\tau)
+G^{\theta\theta}_{j\delta(\eta_1-\eta_1)}(0,0,\tau))}\nonumber\\
&&\times\int_{-\infty}^{+\infty}d\tau_1\partial_x\left(G^{\phi\phi}_{c+(\eta\eta_1)}(x,0,\tau_1)
-G^{\phi\phi}_{c+(\eta-\eta_1)}(x,0,\tau_1)+r_1G^{\phi\theta}_{c+(\eta\eta_1)}(x,0,\tau_1)
-r_1G^{\phi\theta}_{c+(\eta-\eta_1)}(x,0,\tau_1)\right)
\nonumber\\
&&\times\int_{-\infty}^{+\infty}d\tau_2\partial_{x'}\left
(G^{\phi\phi}_{c+(-\eta\eta_1)}(x',0,\tau_2)-G^{\phi\phi}_{c+(-\eta-\eta_1)}(x',0,\tau_2)
+r_1G^{\phi\theta}_{c+(-\eta\eta_1)}(x',0,\tau_2)
-r_1G^{\phi\theta}_{c+(-\eta-\eta_1)}(x',0,\tau_2)\right)
~,\nonumber\\
\label{general_nanotube_noise}
\end{eqnarray}
\end{widetext}
where $g_{(\eta\mu)}$ is the Keldysh Green's function for the STM tip given by Eq.~(\ref{green}), and $\tilde G^{\phi\phi}_{j\delta(\eta\mu)}$ is the Keldysh Green's function for the nanotube bosonic field $\phi_{j\delta}$:
\begin{eqnarray}
&&\tilde G^{\phi\phi}_{j\delta(\eta\mu)}
(x,x^\prime,t-t^\prime)=\bigl\langle T_\mathrm{K}\bigl\{
\phi_{j\delta}(x,t^\eta)\phi_{j\delta}(x^\prime,
t^{\prime\mu})\bigr\}\bigr\rangle\nonumber\\
&&-\frac12\langle
\phi^2(x,t)\rangle-\frac12\langle\phi^2(x^\prime,t^\prime)\rangle~.
\end{eqnarray}

Similar definitions hold for other
combinations of bosonic fields: $\tilde G^{\phi\theta}_{j\delta(\eta\mu)}$, $\tilde G^{\theta\phi}_{j\delta(\eta\mu)}$ and $\tilde G^{\theta\theta}_{j\delta(\eta\mu)}$. For $x=x'=0$, we have\cite{lebedev}:
\begin{eqnarray}\label{green_nano}
      &&\tilde G^{\phi\phi}_{j\delta(-+)}(0,0,t)=-\frac1{2\pi
      K_{j\delta}}\Biggl\{
      \ln\Bigl(1+\frac{iv_{\scriptscriptstyle\rm F}t}a
      \Bigr)\nonumber
      \\
      &&+\sum\limits_{r=\pm1}\sum\limits_{n=1}^\infty
      b_{j\delta}^n\ln\Bigl[1+\frac{iv_{\scriptscriptstyle\rm
      F}t}{a+irnK_{j\delta}L}
      \Bigr]\Biggr\}~,
\end{eqnarray}
where $b_{j\delta}=(K_{j\delta}-1)/(K_{j\delta}+1)$ is the reflection coefficient at the lead positions $x=\pm L/2$. In addition, $\tilde G_{j\delta}^{\phi\theta}(0,0,t)=\tilde G_{j\delta}^{\theta\phi}(0,0,t)=0$, and the Green's function $\tilde G^{\theta\theta}_{j\delta}(0,0,t)$ can be obtained by
the substitution $K_{j\delta}\rightarrow
K_{j\delta}^{-1}$ in Eq.~(\ref{green_nano}).

The integration over $\tau'$ in Eq.~(\ref{courant}) and over $\tau_1$ and $\tau_2$ in Eq.~(\ref{bruit}) can be performed. We obtain:

\begin{widetext}

\begin{eqnarray}\label{current}
\langle I(x)\rangle&=&\frac{4\Gamma^2 e}{(\pi a)^2}sgn(x)\int_{0}^{+\infty}d\tau\frac{\sin(\omega_0\tau)}{\big(1+\left(\frac{v_F\tau}{a}\right)^2\big)^{\frac{1+\nu}{2}}}
\nonumber\\
&\times& \frac{\sin\left((1+\nu)\arctan\big(\frac{v_F\tau}{a}\big)+\frac{1}{8}\sum_{n=1}^{\infty}\left(\frac{b_{c+}^n}{K_{c+}}+(-b_{c+})^n K_{c+}\right)\arctan\left(\frac{2av_F \tau}{a^2+(nLK_{c+})^2-(v_F\tau)^2}\right)\right)}{\prod_{n=1}^{\infty}\left(\left(\frac{a^2+(nLK_{c+})^2-(v_F \tau)^2}{a^2+(nLK_{c+})^2}\right)^2+\left(\frac{2av_F \tau}{a^2+(nLK_{c+})^2}\right)^2\right)^{\frac{1}{16}\left(\frac{b_{c+}^n}{K_{c+}}+(-b_{c+})^n K_{c+}\right)}}~,\\
S(x,x';\Omega=0)&=&\frac{2\Gamma^2 e^2}{(\pi a)^2}\left(1+sgn(x)sgn(x')\right)
 \int_{0}^{+\infty}d\tau\frac{\cos(\omega_0\tau)}{\left(1+\left(\frac{v_F\tau}{a}\right)^2\right)^{\frac{1+\nu}{2}}}
\nonumber\\\label{noise}
&\times& \frac{\cos\left((1+\nu)\arctan\left(\frac{v_F\tau}{a}\right)+\frac{1}{8}\sum_{n=1}^{\infty}\left(\frac{b_{c+}^n}{K_{c+}}+(-b_{c+})^n K_{c+}\right)\arctan\left(\frac{2av_F \tau}{a^2+(nLK_{c+})^2-(v_F\tau)^2}\right)\right)}{\prod_{n=1}^{\infty}\left(\left(\frac{a^2+(nLK_{c+})^2-(v_F \tau)^2}{a^2+(nLK_{c+})^2}\right)^2+\left(\frac{2av_F \tau}{a^2+(nLK_{c+})^2}\right)^2\right)^{\frac{1}{16}\left(\frac{b_{c+}^n}{K_{c+}}+(-b_{c+})^n K_{c+}\right)}}~,
\end{eqnarray}
\end{widetext}
where $\nu=\sum_{j\delta}(K_{j\delta}+1/K_{j\delta})/8$. From Eq.~(\ref{noise}), we immediately see that the zero-frequency cross-correlations $S(x,-x;\Omega=0)$ cancel at order $\Gamma^2$, as it is the case for a non-interacting three terminals device\cite{martin_landauer_buttiker}. This is because quasiparticle excitations in the nanotube suffer multiple reflections at the contacts, 
which lead to a recombination of these in the form of an electron entering the contact on either side, but not both. 

\begin{figure}[h]
\epsfxsize 8 cm
\centerline{\epsffile{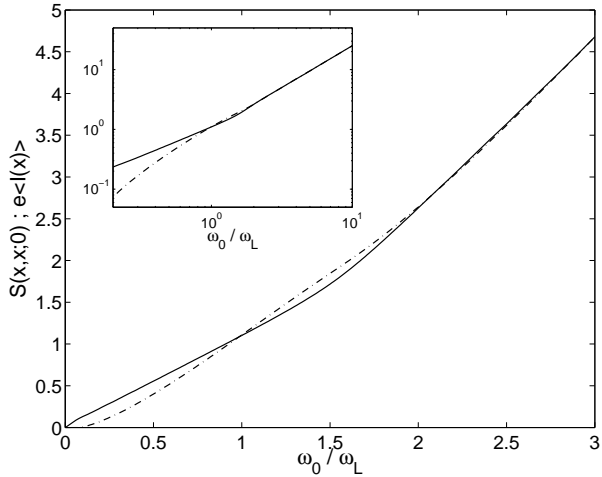}}
\caption{Current (dashed-dotted line) and shot noise (full line) as a function of $\omega_0/\omega_L$, for $K_{c+}=0.2$ and $\omega_c/\omega_L=100$. In the limit $\omega_0/\omega_L\gg 1$, the current and the noise obey the same power law: $(\omega_0/\omega_L)^\nu$ with $\nu=1.4$, whereas for $\omega_0/\omega_L\ll 1$, the current and noise are not more proportional. The inset shows the log-log variations of current and noise.\label{fig2}} 
\end{figure}

By a careful analysis of Eqs.~(\ref{current}) and (\ref{noise}), we notice, as announced in the introduction, that three characteristic frequencies are involved : the voltage frequency $\omega_0=eV_0/\hbar$, the nanotube length frequency $\omega_L=2v_F/K_{c+}L$, which is related to the time $\tau_L=K_{c+}L/2v_F$ that excitations take to reach the leads from the position $x=0$, and the cutoff frequency $\omega_c=v_F/a\gg\omega_0,\omega_L$ of the Luttinger liquid model. Taking realistic values: $a\approx 1 nm$ (carbon nanotube diameter), $L=10 \mu m$ (carbon nanotube length) and $v_F\approx 10^6 m.s^{-1}$ for the Fermi velocity\cite{lemay}, we obtain the estimations $\hbar\omega_c=1 eV$ and $\hbar \omega_L\approx 10^{-3} eV$. We consider that $\hbar\omega_0$ can vary from 0 to $10\hbar\omega_L\approx10^{-2}eV$ (i.e., from 0 to $\approx 10mV$).

In Fig.~\ref{fig2}, numerical calculations of the current and the noise of Eqs. (\ref{current})
and (\ref{noise}) are presented as a function of the ratio $\omega_0/\omega_L$. These are plotted in units of $2e^2\Gamma^2a^{\nu-1}/\pi{\bf \Gamma}(\nu+1)v_F^{\nu+1}$ where $\bf \Gamma$ is the Gamma function. We observe two different regimes. The first one corresponds to the limit $\omega_0/\omega_L\gg 1$ for which the spatial extension of the electron wave packet injected in the nanotube is much smaller than the length of the carbon nanotube: $\Delta x=v_{c+}/\omega_0\ll L/2$. In this limit, the integrals which appear in Eqs. (\ref{courant}) and (\ref{noise}) can be performed exactly\cite{lebedev}, and the current and the noise obey the same power law:  
$S(x,x;\Omega=0)\sim \langle I(x)\rangle \sim (\omega_0/\omega_L)^\nu$, in very good agreement with our numerical results (see inset of Fig.~\ref{fig2}). Therefore, the Schottky relation\cite{schottky} with the electronic charge applies: $S(x,x;\Omega=0)=e|\langle I(x)\rangle|$. As the spatial extension of the electron wave packet is much smaller than the nanotube length, the reflections by the contacts do not play an important role for $\omega_0/\omega_L\gg 1$ (the limit $L\rightarrow +\infty$ is formally equivalent to the limit $b_{c+}\rightarrow0$ where $b_{c+}$ is the reflection coefficient). 

\begin{figure}[h]
\epsfxsize 8 cm
\centerline{\epsffile{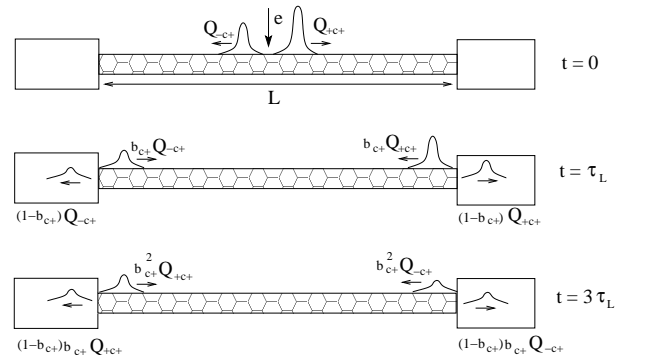}}
\caption{Andreev-like reflections at the contacts between the nanotube and leads. When a wave packet meets a contact (at $t$ equals to a multiple of $\tau_L=\omega_L^{-1}$), part of the charge $Q_{\pm c+}=e(1\pm K_{c+})/2$, equals to $b_{c+}Q_{\pm c+}$, is reflected whereas the other part, equals to $(1-b_{c+})Q_{\pm c+}$, is transmitted into the lead.\label{fig3}}
\end{figure}

A completely different behavior is observed in the limit $\omega_0/\omega_L\le 1$, for which the wave packet spatial extension of the charge excitations is larger, or comparable to the carbon nanotube length: $\Delta x=v_{c+}/\omega_0\ge L/2$. In this regime, the wave packet is split into localized quasiparticle excitations which propagate to the right and to the left contacts, subsequently undergoes Andreev-like reflections, as depicted in Fig.~\ref{fig3}. As a consequence, current and noise have distinct behaviors and are not more proportional to each other. The effect of the reflections by the contacts at the nanotube extremities is to attenuate the Coulomb interactions effect for the noise: from a power law behavior symptomatic of a zero bias anomaly, we end up here with a quasi-linear noise behavior when either the nanotube length is reduced or, equivalently when the voltage decreases. This quasi-linear variation of the noise is extracted form the log-log plot shown in the inset of Fig.~\ref{fig2}. Notice that this quasi-linear variation is shown on less than a decade because for lower values of $\omega_0/\omega_L$, the numerically estimated values of the current and noise become of the order of the numerical error. For this reason, we are not able to charaterize this power law behavior over several decades, as would be suitable in principle.

\begin{figure}[h]
\epsfxsize 8 cm
\centerline{\epsffile{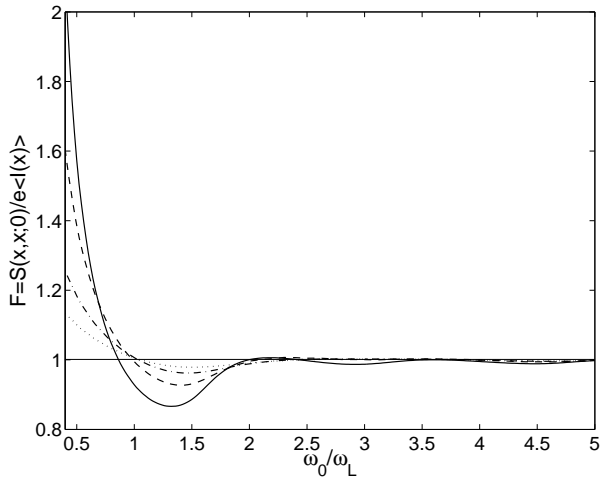}}
\caption{Fano factor as a function of $\omega_0/\omega_L$ for $K_{c+}=0.15$ (full line), $K_{c+}=0.2$ (dashed line), $K_{c+}=0.3$ (dashed-dotted line) and $K_{c+}=0.4$ (dotted line). We take $\omega_c/\omega_L=100$.\label{fig4}}
\end{figure}

The fact that current and noise are not more proportional each other with a coefficient simply equal to $e$ is a novel and interesting feature. In order to further characterize this effect, we have plotted on Fig.~\ref{fig4} the Fano factor, defined by $F=S(x,x;\Omega=0)/e|\langle I(x)\rangle|$, as a function of $\omega_0/\omega_L$, for several values of the Coulomb interactions parameter $K_{c+}$ which differ from the non-interacting case. We observe three distinct phases: super-poissonian noise (i.e., $F>1$) for $\omega_0/\omega_L<1$, sub-poissonian noise (i.e., $F<1$) for $1<\omega_0/\omega_L<2$  and poissonian noise (i.e., $F\approx 1$) for $\omega_0/\omega_L>2$. Non-poissonian noise is observed in the regime where the Andreev-type reflections by the contacts play an important role (i.e. $\omega_0/\omega_L\approx 1$). We conclude that finite size effects are responsible for this non-poissonian character.

In addition, Fig.~\ref{fig4} shows that the Fano factor converges to the value 1 when $K_{c+}$ increases from 0.15 to 0.4. This is fully consistent with the value $F=1$ that is obtained in the absence of Coulomb interactions (where $K_{c+}=1$). The fact that in carbon nanotubes, the Fano factor exhibits a voltage dependence has recently been obtained in some regimes, both experimentally and theoretically\cite{thielmann,onac,recher}. In the present work, the voltage dependent Fano factor allows to test the strength of electronic 
correlations in the nanotube. As shown in the plot of Fig.~\ref{fig4}, to a 
good approximation the width $W$ of the sub-poissonian phase is equal to $\omega_0/\omega_L$.
It can therefore be used for an experimental determination of the Luttinger liquid Coulomb interactions parameter as $K_{c+}=2\hbar v_F W/eV_0L$.

\vspace{0.5cm}

%
%
%
%

\section{photo-assisted shot noise}

We now turn to the response of the system to a voltage AC modulation superposed to the constant DC voltage: $V(t)=V_0+V_1\cos(\omega t)$. 
With this voltage, the tunnel amplitude becomes:
\begin{eqnarray}
\Gamma(t)&=&\Gamma\exp\left(i\omega_0 t+i\frac{\omega_1}{\omega}\sin(\omega t)\right)\nonumber\\
&=&\Gamma\sum_{p=-\infty}^{+\infty}J_p\left(\frac{\omega_1}{\omega}\right)\exp(i(\omega_0+p\omega)t)~,
\end{eqnarray}
where $J_p$ the Bessel function on order $p$ (even or odd integer) and $\omega_1=eV_1/\hbar$. 
Calculations of current and zero-frequency noise are similar to the DC 
voltage case, except that we have now an infinite sum of Bessel functions\cite{platero}. 
The calculation is thus analogous to the one which applies to 
the fractional quantum Hall effect\cite{crepieux2}, except that here 
only electrons tunnel in the nanotube.
We thus obtain:
\begin{widetext}
\begin{eqnarray}\label{AC_current}
\langle I(x)\rangle&=&\frac{4\Gamma^2 e^2}{(\pi a)^2}sgn(x) \sum_{p=-\infty}^{+\infty}J^2_p\left(\frac{\omega_1}{\omega}\right)\int_{0}^{+\infty}d\tau\frac{\sin((\omega_0+p\omega)\tau)}{\left(1+\left(\frac{v_F\tau}{a}\right)^2\right)^{\frac{1+\nu}{2}}}\nonumber\\
&\times& \frac{\sin\left((1+\nu)\arctan\left(\frac{v_F\tau}{a}\right)+\frac{1}{8}\sum_{n=1}^{\infty}\left(\frac{b_{c+}^n}{K_{c+}}+(-b_{c+})^n K_{c+}\right)\arctan\left(\frac{2av_F \tau}{a^2+(nLK_{c+})^2-(v_F\tau)^2}\right)\right)}{\prod_{n=1}^{\infty}\left(\left(\frac{a^2+(nLK_{c+})^2-(v_F \tau)^2}{a^2+(nLK_{c+})^2}\right)^2+\left(\frac{2av_F \tau}{a^2+(nLK_{c+})^2}\right)^2\right)^{\frac{1}{16}\left(\frac{b_{c+}^n}{K_{c+}}+(-b_{c+})^n K_{c+}\right)}}~,\\\label{AC_noise}
S(x,x';\Omega=0)&=&\frac{2\Gamma^2 e^2}{(\pi a)^2}\left(1+sgn(x)sgn(x')\right) \sum_{p=-\infty}^{+\infty}J^2_p\left(\frac{\omega_1}{\omega}\right)\int_{0}^{+\infty}d\tau\frac{\cos((\omega_0+p\omega)\tau)}{\left(1+\left(\frac{v_F\tau}{a}\right)^2\right)^{\frac{1+\nu}{2}}}\nonumber\\
&\times& \frac{\cos\left((1+\nu)\arctan\left(\frac{v_F\tau}{a}\right)+\frac{1}{8}\sum_{n=1}^{\infty}\left(\frac{b_{c+}^n}{K_{c+}}+(-b_{c+})^n K_{c+}\right)\arctan\left(\frac{2av_F \tau}{a^2+(nLK_{c+})^2-(v_F\tau)^2}\right)\right)}{\prod_{n=1}^{\infty}\left(\left(\frac{a^2+(nLK_{c+})^2-(v_F \tau)^2}{a^2+(nLK_{c+})^2}\right)^2+\left(\frac{2av_F \tau}{a^2+(nLK_{c+})^2}\right)^2\right)^{\frac{1}{16}\left(\frac{b_{c+}^n}{K_{c+}}+(-b_{c+})^n K_{c+}\right)}}~.
\end{eqnarray}
\end{widetext}
In Sec. III, we showed that the integrals which appear in Eqs.~(\ref{AC_current}) and (\ref{AC_noise}) behave like a power law $\propto \omega_0^\nu$ when $\omega_0\gg\omega_L$.Thus, in the presence of an AC voltage modulation, the current and zero-frequency noise give an infinite sum of power laws, 
which cannot be understood straightforwardly as in the preceding section. 

\begin{figure}[h]
\epsfxsize 8cm
\centerline{\epsffile{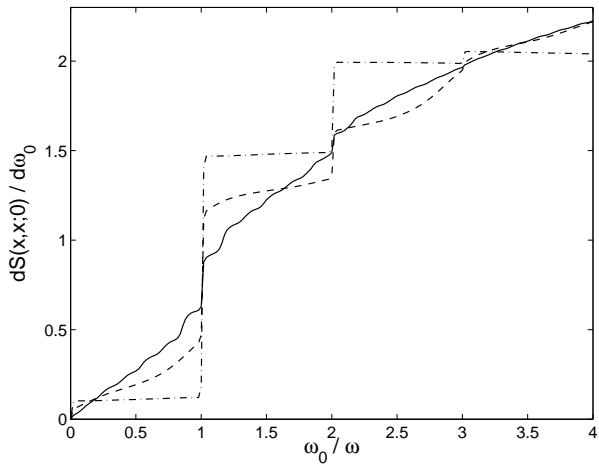}}
\caption{Noise derivative as a function of $\omega_0/\omega$ for $K_{c+}=0.2$, $\omega_1/\omega=2$, $\omega_c/\omega=100$ and different values of the nanotube length frequency: $\omega_L/\omega=0.1$ (full line), $\omega_L/\omega=1.2$ (dashed line) and $\omega_L/\omega=6$ (dashed-dotted line).\label{fig5}}
\end{figure}

The ``standard'' way\cite{lesovik} to display the results for 
photo-assisted transport is to consider the noise derivative as a function 
of voltage: in particular, this allows to compare the results with the 
non-interacting case where the noise derivative exhibits a staircase variation, which corresponds to the normal metal junction of Ref.~\onlinecite{lesovik} for a Fermi liquid. In Fig.~\ref{fig5}, we plot the numerically computed noise derivative as a function of the ratio $\omega_0/\omega$ in the presence of Coulomb interactions ($K_{c+}=0.2$). We see different kinds of behaviors, which again have to do with the finite size effects in the nanotube. 

For $\omega_L/\omega=0.1$ (full line), we are in the limit where the wave packet spatial extension is smaller that the nanotube length. We therefore invoke the same arguments as before, concerning the recombination of electron wave packets
into simple electrons: in this regime (except for a small region close to the origin)  
the vast majority of the voltage scale lies in the regime where $\omega_0>\omega_L$. 
The noise derivative differs from the single electron 
behavior, in the sense that the sharp steps and plateaus expected in this case are absent. 
Instead, because of Coulomb interactions effect, the noise derivative is smoothed out, but there is a clear reminiscence
of the step positions: the slope of $dS(x,x;\Omega=0)/d\omega_0$ increases
abruptly at the location of these steps. We attribute the smoothing to the 
tunneling density of states on the nanotube which is modified by the Coulomb interactions in the nanotube. Between the steps, we observe oscillations which are originate from the commensurability between the nanotube length frequency and the AC frequency.

For $\omega_L/\omega=1.2$ (dashed line), we are in an intermediate regime for which electron wave packets are comparable to the nanotube length. For $\omega_L/\omega=6$ (dashed-dotted line), we are in the limit where electron wave packets are larger than the nanotube length, and as a consequence, the finite size effects dominate over the Coulomb interactions effect and a stepwise behavior in $dS(x,x;\Omega=0)/d\omega_0$, which is typical of non-interacting metals, can be identified. 

Finally, we notice that for $\omega_0\gg\omega_L$, all the curves converge to the simple expression for the noise derivative given by the sum $\sum_{p=-\infty}^{+\infty}J^2_p(\omega_1/\omega)sgn(\omega_0+p\omega)|\omega_0+p\omega|^{\nu-1}$. The reason is the following one: in this limit, Coulomb interactions play an important role and the DC noise has a power law dependency with the voltage (as shown in section III): $S_{DC}(\omega_0)\propto |\omega_0|^\nu$ which lead to a AC noise of the form:
\begin{eqnarray}
S_{AC}(\omega_0)&=&\sum_{p=-\infty}^{+\infty}J^2_p\left(\frac{\omega_1}{\omega}\right)S_{DC}(\omega_0+p\omega)\nonumber\\
&\propto& \sum_{p=-\infty}^{+\infty}J^2_p\left(\frac{\omega_1}{\omega}\right)|\omega_0+p\omega|^\nu~,
\end{eqnarray} 

where the exponent $\nu$ is related to the Coulomb interactions parameter.

To summarize, Coulomb interactions affect the height and shape of the steps in the differential noise, and trigger oscillations between these steps when $\omega_L/\omega$ is small.

%
%
%
%

\section{Conclusion}   

This study has dealt with the finite size effects for electron
injection on a nanotube connected to Fermi liquid leads, from the point of view 
of zero-frequency noise as well as for photo-assisted transport. 
On the first topic, we have shown that at low and intermediate 
voltages compared to the frequency length scale, the current and noise 
deviate strongly from the Schottky relation, leading first to super-poissonian noise, 
then to sub-poissonian behavior as compared to tunneling in non-interacting system. 
The sub-poissonian result can be exploited toward an experimental 
determination of the Luttinger liquid interactions parameter in 
the nanotube.

A qualitative understanding of the deviation from the Schottky 
behavior can be reached by arguing that in the limit
$\omega_0/\omega_L\gg 1$, the quasi-particle wave packets which originate from the 
injected electron are well localized in space: their spatial extension is much smaller that the nanotube length and a recombination as an electron in either lead is possible. 
On the opposite, $\omega_0/\omega_L\ll 1$ ``wide'' wave packets
have not enough space on the nanotube length, and the Fabry-Perot\cite{safi_maslov}
process for transforming quasiparticle excitations into electrons at the leads is simply not efficient. Granted, it is however difficult to explain in detail the 
transition from super-poissonian noise to sub-poissonian noise: this qualitative argument 
can so far only predict a deviation from the electron charge in the Schottky formula.   

For photo-assisted transport, generally speaking as stepwise structure 
is identified, which is similar to that of single electrons, but 
steps are smoothed out due to the fact that electrons tunnel 
in a strongly correlated one dimensional system, where electrons are 
not truly welcome.
More interestingly, we identify that when the width of the electron 
wave packets becomes comparable or exceeds the nanotube length, 
a stepwise behavior is observed. Not surprisingly, 
this happens precisely in the same regime where the deviation from Schottky 
behavior was observed.

Note that in contrast to our previous work dedicated to the detection 
of quasi-particle anomalous charges  in a nanotube connected to 
leads\cite{lebedev}, here we can make a diagnosis on finite size 
effects in transport by a relatively simple zero-frequency 
measurement (by zero-frequency, in experiments we really mean
frequencies which are sufficiently low so that $1/f$ noise
does not dominate). This is a rather important practical aspect 
of this work: indeed, for nanotubes length of the order of 
$1\mu m$, the frequency scales needed for the finite-frequency noise
cross correlations diagnostic of Ref.~\onlinecite{lebedev}
are of the order of $100 GHz - 1 THz$. Such frequencies are difficult to 
detect by conventional techniques, and require an 
on-chip noise measuring apparatus\cite{onac,billangeon}. The fact that a low frequency 
auto-correlation noise measurement is sufficient to probe 
finite size effects is clearly an advantage here.    

The present results could be tested experimentally, using essentially the same
geometry as in the experiments of Ref.~\onlinecite{dekker}. There, electrons 
were injected by an STM tip on a suspended nanotube, which was placed across 
a trench composed of highly doped silicon. Both sides of the trench were 
short circuited and connected to the ground.
Because the contacts were of bad quality, 
the physics observed in this experiment was that of the Coulomb blockade, with 
additional features in the current voltage characteristics due to the lateral 
breathing vibration mode. For the purposes of our present findings, 
only one aspect of this experiment should be modified: the quality of the 
contacts should be improved to achieve a rather high transparency. 
The fact that the Dekker experiment uses a short-circuited trench 
is not detrimental for our consideration because the auto-correlation 
noise provides us with all the information we need, and according to 
our results the noise measured on either side of the STM tip is the same.

Upon completion of this manuscript, we noted a preprint \cite{orellana} which deals
with photo-assisted transport in carbon nanotubes. However, this study does
not take into account the effect of Coulomb interactions, which is the main focus
of our work, and it is confined to the calculation of the conductance as opposed
to our noise calculation.

T.M. and A.C. acknowledge support by an ``Action Concert\'ee Nanosciences'' from CNRS.

%
%
%
%

\end{document}